\documentclass[12pt]{iopart}

\newcommand{\gdot}{\dot{\gamma}}
\usepackage[dvips]{graphicx}
\usepackage[english]{babel}
\usepackage{epsfig}

\begin{document}
\bibliographystyle{apsrev}

\title{Dynamic Response of Block Copolymer Wormlike Micelles to Shear Flow.}
\author{B. Lonetti, J. Kohlbrecher\dag, L. Willner, J. K. G. Dhont, and M. P. Lettinga} \address{Institut f\"ur Festk\"orperforschung,
Forschungszentrum J\"ulich, D-52425 J\"ulich, Germany}
\dag\address{Laboratory for Neutron Scattering, ETH Z\"{u}rich and Paul Scherrer Institut, 5232 Villigen PSI, Switzerland}
\date{\today}
\newcommand{\dg}{\mbox{$\dot{\gamma}$}}
\newcommand{\uu}{\mbox{$\hat{\mathbf{u}}$}}
\newcommand{\nn}{\mbox{$\hat{\mathbf{n}}$}}
\newcommand{\rr}{\mbox{$\mathbf{r}$}}
\newcommand{\rot}{\mbox{$\hat{\mathcal{R}}$}}

\begin{abstract}

The linear and non-linear dynamic response to an oscillatory shear
flow of giant wormlike micelles consisting of Pb-Peo block
copolymers is studied by means of Fourier transform rheology.
Experiments are performed in the vicinity of the isotropic-nematic
phase transition concentration, where the location of
isotropic-nematic phase transition lines is determined
independently. Strong shear-thinning behaviour is observed due to
critical slowing down of orientational diffusion as a result of
the vicinity of the isotropic-nematic spinodal. This severe
shear-thinning behaviour is shown to result in gradient shear
banding. Time-resolved Small angle neutron scattering experiments are used to
obtain insight in the microscopic phenomena that underly the
observed rheological response. An equation of motion for the
order-parameter tensor and an expression of the stress tensor in
terms of the order-parameter tensor are used to interpret the experimental data, both
in the linear and non-linear regime. Scaling of the dynamic
behaviour of the orientational order parameter and the stress is
found when critical slowing down due to the vicinity of the
isotropic-nematic spinodal is accounted for.
\end{abstract}


\maketitle

\section{Introduction}

Dispersions of surfactant wormlike micelles form a class of
systems that has been intensively studied during the past two
decades. Wormlike micellar systems sometimes exhibit extreme
shear-thinning behaviour\cite{Berret04}, resulting in
shear-induced structure formation like shear
banding\cite{Manneville08}. Strong shear thinning is of practical
interest, since often systems are required in practical
applications that exhibit extreme differences in viscosity between
the sheared and quiescent state. The reason for the popularity of
wormlike micelles lies in their complex rheological behavior like
shear banding and chaotic response, which are connected to the
thinning behaviour of these systems\cite{Berret04}. It is
therefore important to understand the microscopic mechanism
underlying the very strong shear-thinning behaviour of wormlike
micelles. There are several possible microscopic mechanisms that
could be responsible for the occurrence of strong shear
thinning\cite{Cates90a}. One mechanism is related to the breaking
and/or merging of worms. Scission due to shearing forces and
merging of worms through stressed entanglement
points\cite{Briels04} can lead to strong shear thinning. Another
possible mechanism for strong shear thinning is connected to the
fact that wormlike systems can undergo an isotropic-nematic (I-N)
phase transition. Rotational diffusion close to I-N spinodal lines
in the phase diagram is very slow, so that a relatively strong
alignment on applying shear flow occurs. Such a strong increase in
the degree of alignment leads in turn to strong shear thinning. We
shall hereafter refer to the slowing down of rotational diffusion
close to the I-N spinodals simply as "{\it critical slowing
down}". By definition, the rotational diffusion coefficient at the
spinodal changes sign, and is therefore zero at the spinodal,
which implies very slow rotational Brownian motion.

For most studied surfactant wormlike micellar systems, the I-N
transition occurs at relatively high volume fractions of around
$10\; \%$. At this high concentration the viscosity of the system
is quite large, and moreover, a transition to a gel phase can
interfere. For CPLC/NaSal in brine, for example, gelling occurs in the vicinity
of the I-N transition on changing the temperature by just a few
degrees. Furthermore, the I-N transition is only found under flow
conditions. These features complicate detailed studies on the rheological response
of wormlike micelles and its microscopic origin. We therefore
study here a system that exhibits many of the properties of surfactant
wormlike micellar systems that are responsible for their
interesting rheological behaviour, but that does not have the
above mentioned complications. For an
I-N transition to occur without flow, we need a system where the
persistence length $l_{p}$ is much larger than the thickness $d$
of the chains. The ratio $l_p/d$ should be larger as compared to
typical values for wormlike micelles. A candidate system could be
micelles formed from block-copolymers. A well studied system is
poly(butadiene)-poly(ethylene oxide) (Pb-Peo) diblock copolymer
with a $50-50$ block composition in aqueous solution. The main advantage of this
system is that it is very stiff, with a persistence length of
around $500\;nm$ and a diameter of $14\;nm$.
The contour length of the Pb-Peo worms is around $1\;\mu m$.
As a result the large ratio $l_{p}/d$ as compared to common surfactant micellar systems,
the diblock copolymer system shows an I-N transition at a modest concentration of about
$5\; \%$, although the transition concentration has not been
determined accurately yet\cite{Won99}. Other advantages of the Pb-Peo
system are that it is possible to tune the monomer-exchange kinetics
between the polymers\cite{Lund06} or its morphology\cite{Denkova08} by using different solvent
mixtures. Furthermore, these polymers are easily
marked with fluorescent dyes, which enables their visualization
with fluorescence microscopy. In a recent study F\"{o}rster et al.
used this system, amongst others, for Rheo-SANS measurements,
where stationary shear measurements were combined with Small Angle
Neutron Scattering (SANS)\cite{Foerster05}. A feature of this
diblock copolymer system that is probably not shared with micellar
systems is that the polymers do not easily break and merge under
flow. We thus focus on the microscopic mechanism mentioned above,
related to critical slowing down of rotational diffusion close to
the I-N transition.

In section \ref{Theory}, a well-known theoretical frame work for
the dynamics and rheological behaviour of stiff rods is
summarized. This theory does not include flexibility of single
polymer chains, but does include the slowing down of rotational
diffusion due to the vicinity of the I-N spinodal. This theory
will be used to assess the effect of the vicinity if the I-N
transition on rheological response. A comparison of our
experiments with predictions based on this theory can only be done
on a qualitative level, since flexibility is neglected in the
theory. After the materials section we describe a newly developed
time-resolved SANS set-up, and the couette cells and Rheometers
that were used. In section \ref{results} we first discuss the flow
curve of the system and determine the corresponding flow profiles.
It is also shown in this section how the (non-equilibrium) binodal
line can be found from shear step-down experiments. Then we
discuss SANS experiments on quiescent and stationary sheared
systems, which we need as an input in the last subsection on
dynamic experiments. In the latter subsection we connect the
time-resolved SANS measurements with Fourier Transform Rheology
results. The spinodal point is determined in order to establish
whether the concept of critical slowing down indeed applies.

\section{Theory}\label{Theory}
\subsection{Concentration dependence of the rotational diffusion coefficient}

On approach of the isotropic-nematic (I-N) spinodal, the {\it
collective} rotational diffusion coefficient vanishes and becomes
negative in the unstable part of the phase diagram. As will be
discussed later, this rotational diffusion coefficient describes
the dynamics of small perturbations of the orientational order
parameter from its value in a stationary state. For a system of
very long and thin, rigid rods with repulsive interactions that
have a range that is small compared to the length of the rods,
critical slowing is described by the equation of motion for the
orientational order parameter tensor
$\mathbf{S}\equiv<\hat{\mathbf{u}}\hat{\mathbf{u}}>$, where
$\hat{\mathbf{u}}$ is the unit vector along the long axis of a
rod, which specifies the orientation of the rod, and where the
brackets indicate ensemble averaging. Starting from the
Smoluchowski equation for rod-like colloids with hard-core
interactions, an equation of motion for $\mathbf{S}$ can be
derived\cite{Dhont03c}, that is similar to the Doi-Edwards
equation of motion\cite{Doi86},
\begin{eqnarray} \label{eq_St}
\frac{d\;}{dt}\mathbf{S}= -6D_{r}\left\{ \mathbf{S}-\mbox{\small{$\frac{1}{3}$}}\hat{\mathbf{I}}
+\mbox{\small{$\frac{L}{D}$}}\varphi\,\left( \mathbf{S}^{(4)}:\mathbf{S}-\mathbf{S}\cdot\mathbf{S} \right)\right\}
+\dg \left\{ \hat{\mathbf{\Gamma}}\cdot\mathbf{S}+\mathbf{S}\cdot\hat{\mathbf{\Gamma}}^{T}\!
-2\mathbf{S}^{(4)}:\hat{\mathbf{E}}\right\} ,
\end{eqnarray}
where $D_{r}$ is the rotational diffusion coefficient at infinite
dilution, $L$ is the length of the rods, $d$ their core diameter,
$\varphi$ is the volume fraction of rods, $\dot{\gamma}$ is the
shear rate, $\hat{\mathbf{\Gamma}}$ is the velocity-gradient
tensor and
$\hat{\mathbf{E}}=\mbox{\small{$\frac{1}{2}$}}[\hat{\mathbf{\Gamma}}+\hat{\mathbf{\Gamma}}^{T}]$
is the symmetrized velocity-gradient tensor (where the superscript
"$T\,$" stands for "transpose"). Furthermore,
$\mathbf{S}^{(4)}\equiv
<\hat{\mathbf{u}}\hat{\mathbf{u}}\hat{\mathbf{u}}\hat{\mathbf{u}}>$
is a fourth-order tensor. A closure relation that expresses
contractions of the form $\mathbf{S}^{(4)}:\mathbf{M}$ in terms of
$\mathbf{S}$ for arbitrary second rank tensors $\mathbf{M}$ is
discussed in ref.\cite{Dhont03c},
\begin{eqnarray}\label{closure}
<\uu\,\uu\,\uu\,\uu>:\!\mathbf{M}=
\mbox{\small{$\frac{1}{5}$}}\left\{ \mathbf{S}\!\cdot\!\overline{\mathbf{M}}
+\overline{\mathbf{M}}\!\cdot\!\mathbf{S}
-\mathbf{S}\!\cdot\!\mathbf{S}\!\cdot\!\overline{\mathbf{M}}
-\overline{\mathbf{M}}\!\cdot\!\mathbf{S}\!\cdot\!\mathbf{S}
+2\mathbf{S}\!\cdot\!\overline{\mathbf{M}}\!\cdot\!\mathbf{S}
+3\mathbf{S}\,\mathbf{S}\!:\!\overline{\mathbf{M}}\right\} \,,
\end{eqnarray}
where $\overline{\mathbf{M}}=\mbox{\small{$\frac{1}{2}$}}[ \mathbf{M}+\mathbf{M}^{T}]$ is the symmetric part of the tensor $\mathbf{M}$. For simple shear flow, the velocity-gradient tensor has the form,
\begin{eqnarray}\label{velgrad}
\hat{\mathbf{\Gamma}}\,=\,\left(
\begin{array}{ccc}
0 & 1 & 0 \\
0 & 0 & 0 \\
0 & 0 & 0 \end{array} \right) \,,
\end{eqnarray}
which corresponds to a flow in the $x$-direction with its gradient
in the $y$-direction.

The largest eigenvalue $\lambda$ of $\mathbf{S}$, the
"orientational order parameter", is a measure for the degree of
alignment (for the isotropic state $\lambda=1/3$ and for a
perfectly aligned state, $\lambda=1$). In order to illustrate
critical slowing down of orientational diffusion, we consider
first an isotropic state which is slightly perturbed. The equation
of motion for a small perturbation $\delta\lambda$ of
$\lambda=1/3$ in the isotropic state, in the absence of flow, is
readily obtained from eq.(\ref{eq_St}) (together with the closure
relation (\ref{closure}),
\begin{eqnarray} \label{eq_lambda}
\frac{d\delta\lambda}{dt}\;=\;-6D_r\left\{1-\frac{1}{4}\frac{L}{d}\phi\right\}\delta\lambda\;
=\;-6D_r^{eff}\delta\lambda\,
\end{eqnarray}
where,
\begin{eqnarray} \label{Eq_deff}
D_r^{eff}=D_r\left\{1-\frac{1}{4}\frac{L}{d}\phi\right\}
\end{eqnarray}
is the effective rotational diffusion coefficient. Hence,
\begin{eqnarray} \label{eq_lambdat}
\delta\lambda(t)\;=\delta\lambda(t=0)\;\exp{\{-6\,D_r^{eff}\,t\}}\;.
\end{eqnarray}
From eq.(\ref{Eq_deff}) it can be seen that
$D_{r}^{eff}\rightarrow 0$ as $(L/d)\varphi\rightarrow 4$.
Collective rotational diffusion thus becomes very slow on approach
of the spinodal concentration where $(L/d)\varphi = 4$. For larger
concentrations, where $D_{r}^{eff}<0$, the isotropic state is
unstable, and the initially small orientational order parameter
increases in time. In the presence of shear flow, the above
analysis must be done numerically, since the unperturbed (stable
or unstable) stationary state under shear flow is not known
analytically. The effective rotational diffusion coefficient is
now a tensorial quantity rather than a scalar as for the isotropic
state discussed above. The phenomenon of critical slowing down,
however, is unchanged : rotational diffusion becomes very slow on
approach of the spinodal (where at least one of the eigenvalues of
the rotational diffusion tensor changes sign). This slowing down
of rotational diffusion has pronounced effects on the
shear-thinning behaviour, as will be discussed later.

\subsection{Dynamic response of stress and orientational order}

From microscopic considerations, an expression for the stress
tensor $\mathbf{\Sigma}$ can be obtained\cite{Dhont03b}, which is similar to an earlier
derived expression by Doi and Edwards\cite{Doi86},
\begin{eqnarray}\label{Eq_stress}
\mathbf{\Sigma}_{D}= 2\,\eta_{0}\,\dg\,\mbox{\large{$[$}} \hat{\mathbf{E}} + \frac{\left( L/D \right)^{2}}{3\ln\{L/D\}}\,\varphi
\left\{\hat{\mathbf{\Gamma}}\cdot
\mathbf{S}+\mathbf{S}\cdot\hat{\mathbf{\Gamma}}^{T}-\mathbf{S}^{(4)}:\hat{\mathbf{E}}-\mbox{\small{$\frac{1}{3}$}}\,
\hat{\mathbf{I}}\,\mathbf{S}:\hat{\mathbf{E}}-\frac{1}{\dg} \frac{d\mathbf{S}}{dt}\right\}\mbox{\large{$]$}}\,.
\end{eqnarray}
For an oscillatory shear flow, the shear-rate $\dot{\gamma}$ in
eqs.(\ref{eq_St},\ref{Eq_stress}) is time dependent,
\begin{eqnarray}\label{shearrate}
\dg (t)\;=\;\dg_{0}\,\cos\{\omega\,t\}\;,
\end{eqnarray}
where $\dg_{0} = A\,\omega$ is the shear-amplitude, with $A$ the
strain amplitude and $\omega$ the frequency of oscillation.

The linear and non-linear response of suspensions or rigid rods,
within the approximations involved in the theory, can be obtained
from numerical solutions of eqs.(\ref{eq_St},\ref{Eq_stress})
\cite{Dhont03c}. In particular, dynamic response functions can be
obtained from a Fourier analysis of the time dependence of the
stress tensor after transients have relaxed. For sufficiently
large shear rates, higher order non-linear response functions
come into play. For these higher shear-amplitudes, the time
dependent stress tensor must be Fourier expanded as,
\begin{eqnarray}\label{Eq_FourierSig}
\mathbf{\Sigma}_{D}\;=\;2\,\dot{\gamma}_{0}\,\hat{\mathbf{E}}\,\sum_{n=0}^{\infty}\,
\,|\eta|_{n}\,\sin( n\omega t+\delta_n )\,\;,
\end{eqnarray}
where $|\eta|_{n}$ and $\delta_n$ are the amplitude and phase shift of the Fourier components, respectively. Similarly, the scalar orientational order
parameter will respond in a non-linear fashion, so that,
\begin{equation}\label{Eq_FourierP2}
P_{2}(t)=\sum_n^{\infty}\,|P_2|_{n}\cos(\omega
t+\epsilon_n)\,\;,
\end{equation}
where $P_{2}=\mbox{\small{$\frac{1}{2}$}}[3\lambda -1 ]$ (as
before, $\lambda$ is the largest eigenvalue of $\mathbf{S}$). It
should be mentioned that in scattering experiments only
projections of the orientational order parameter tensor are
probed. In that case, $P_{2}$ in eq.(\ref{Eq_FourierP2}) does not
correspond to the largest eigenvalue of $\mathbf{S}$, but only to
the corresponding projection of $\mathbf{S}$. In the experiments
described in this paper the vorticity-flow plane is probed, for
which it is readily shown  from eq.(\ref{eq_St}) by expanding
$\mathbf{S}$ for small shear rates, that the leading term in shear
rate varies like $\sim \dot{\gamma}^{2}$. The time-dependence of
the experimentally determined orientational order parameter term
has therefore the double frequency of the applied shear flow.

One may ask about the shear rate beyond which non-linear response
is expected, and beyond which a frequency phase shift will be
found. An analysis of the equation of motion (\ref{eq_St}) and the
expression (\ref{Eq_stress}) for the stress tensor for the
isotropic state and to leading order in non-linearity reveals that
the so-called effective Peclet number,
\begin{eqnarray}\label{Eq_Peclet}
Pe_{eff}\;=\;\dot{\gamma}_{0}/D_{r}^{eff}\;,
\end{eqnarray}
and the effective Deborah number,
\begin{eqnarray}\label{Eq_Deborah}
\Omega_{eff}\;=\;\omega /D_{r}^{eff}\;,
\end{eqnarray}
measure the non-linearity and phase shift. Here, the effective
rotational diffusion coefficient is given in eq.(\ref{Eq_deff}).

\section{Material}

In this study we used a symmetric Pb-Peo block copolymer prepared
by living anionic polymerization; the
synthesis follows a two-step procedure since the polymerization
conditions for ethylene oxide are different from those for
butadiene. Details of the two-step procedure can be found in an
earlier publication\cite{Allgaier97}. The Pb-Peo block copolymer was
characterized by size exclusion chromatography (SEC) using a
mixture of tetrahydrofuran/dimethylacetamide 90/10 v/v as eluant.
The polydispersity, Mw/Mn,  of the block copolymer was smaller
than 1.04. No signs of PEO and PB homopolymers were found in the
SEC-chromatograms. Absolute molecular weights were determined by
1H-NMR measurements in CDCl3. Thereby, the signal of the t-butyl
initiator group was taken as an internal reference. The number
average molecular weights, Mn, are 2.6 Kg/mol  for PB and 2.64
Kg/mol for PEO. Polymer solutions were prepared by dissolving the
polymer in D$_{2}$O (Chemotrade,\% D = 99.8 \% ) and, in order to
guarantee its complete dissolution, especially in the case of the high
concentration samples, they were kept for half an hour at
56$^{o}$C and then let cool down slowly to the ambient
temperature. When not specified otherwise, the concentrations will
be expressed as weight fraction.

\section{Experimental}

SANS experiments have been performed at the SANS I instrument at
the SINQ spallation source at the Paul Scherrer Institute (PSI) in
Villigen, Switzerland\cite{Kohlbrecher00}. We used thermal
neutrons of wavelength $\lambda = 0.8\; nm$ with a wavelength
spread $\Delta\lambda/\lambda$ of about 0.1. The data analysis was
performed using the BerSANS software package \cite{Keiderling02}.
A standard water sample was used for calibration of absolute
scattering intensities and also to account for non-uniform
detector efficiency. For the Rheo-SANS experiments a Rheowis
strain controlled Rheometer with a Couette type shear cell (bob:
$48\; mm$ radius, cup: $50\; mm$ radius) was placed in the neutron
beam in the so-called radial configuration. In this configuration
the neutron beam passes through the center of the sapphire cell,
transparent for neutrons, and is parallel to the gradient
direction so that the flow-vorticity plane is probed by the 2D
detector. The accessible torque range is between $10^{-7}$ and
$0.046\;Nm$, the frequency range between $5\times 10^{-3}$and
$10\;Hz$ and the amplitude range between $5\times 10^{-2}$ and
$45°$. Both steady state and oscillatory experiments were
performed. In order to probe the time dependent structural changes
with SANS under oscillating shear, a stroboscopic data acquisition
scheme, implemented on the SANS-1 instrument, has been used. The
electronics of the rheometer supplies a low and high signal
depending on the turning direction. The falling edge of this
rectangular signal has been used to trigger the data acquisition
of the scattered neutrons, producing histograms of $128\;×\;128$
pixels of $0.75\times 0.75\; cm^2$ spatial resolution and at least
$n=100$ time channels of widths $\Delta t=(n\times
\omega/2\pi)^{-1}$  where $\omega/2\pi$  is the frequency of the
applied oscillating shear. The time of flight $t_{tof}$ of the
scattered neutrons between sample and detector has been corrected
to obtain the exact phase between applied shear and scattered
neutrons. However, this correction can be practically neglected as
the applied shear frequencies are much lower than 1/$t_{tof}$.
Before starting the neutron data acquisition the rheometer was
oscillating several cycles to assure that no transient effects
were measured. To obtain sufficient counting statistics for each
time channel, the histograms of many shear cycles were summed up
over a time going from one hour to fifteen minutes for the lowest
and highest concentration respectively. With this technique the
temporal evolution of the structural alignment of the diblock
copolymers during a whole shear cycle could be measured.

Fast Fourier Transform Rheological experiments were performed on a
strain controlled rheometer (ARES, TA instruments), using a
couette geometry (bob: $32\; mm$ radius, cup: $34\; mm$ radius).
The stress response to dynamic strain experiments has been
simultaneously recorded with a Analog Digital Card and analyzed
with Fast Fourier Transform software as described in
ref.\cite{Wilhelm98}. The same instrument was used for step-rate
experiments and to obtain flow curves. Spatially resolved velocity
profiles were measured on a homebuild Heterodyne Dynamic Light Scattering set-up using a
closed, transparent Couette cell ($2\; mm$ gap), see e.g. ref.\cite{Salmon03c}.

\begin{figure}\center
\includegraphics[width=1\textwidth]{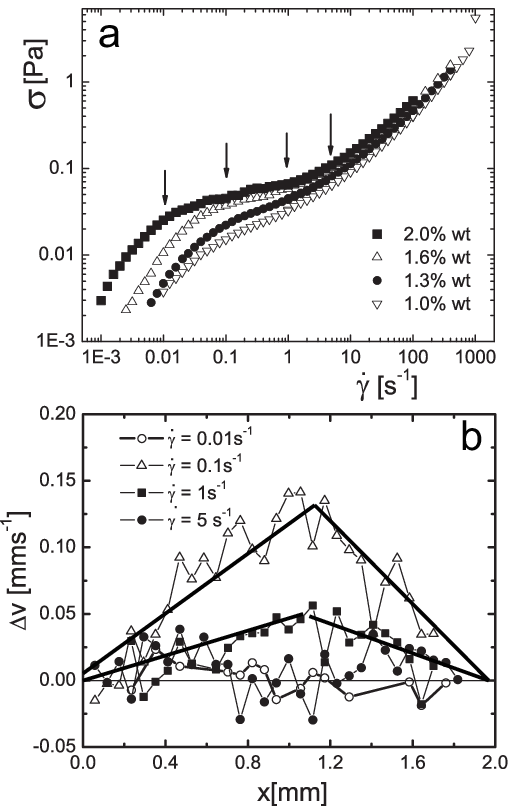}
\caption{(a) Flow curves for different Pb-Peo concentrations. (b)
The relative flow velocity $\Delta v(x)\;=\;V(x)-\gdot x$
throughout the gap of the couette cell for various applied shear
rates as indicated  by the arrows in (a): $0.01$, $0.1$, $1$ and
$5 \;s^{-1}$. The lines indicate two shear bands.}
\label{fig:flowvelocitycurve}
\end{figure}

\section{Results and Discussion}\label{results}

\subsection{Flow curve and step-down rheology}
The Pb-Peo block copolymer under study forms wormlike micelles in
water solution. As molecular worm-like micelles consisting of
surfactant molecules, these giant wormlike micelles show a
pronounced shear thinning behavior.
Fig.\ref{fig:flowvelocitycurve}a shows the stress as a function of
the shear rate for Pb-Peo solutions with volume fraction between
$1\; \%$ and $2\; \%$. These concentrations lie close to the
suggested literature value for the I-N transition\cite{Won99}. All the curves in
Fig.\ref{fig:flowvelocitycurve}a exhibit a shear thinning region
which extends to lower shear rates with increasing volume fraction
of micelles, while the corresponding stress plateau becomes
flatter. For the sample with the highest concentration, i.e.
$[Pb-PEO]=2\; \%$, we tested if the sample shows shear banding, as
is expected for extreme shear thinning samples
\cite{dhont08,olmsted08}. A few typical velocity profiles relative
to the applied shear rate within the gap of the couette cell are
plotted in Fig.\ref{fig:flowvelocitycurve}b, as obtained from
spatially resolved heterodyne light scattering measurements. Shear
banding is observed between $0.1 \;s^{-1}$ and $0.75\; s^{-1}$,
which corresponds with the flat region in the flow curve in
Fig.\ref{fig:flowvelocitycurve}a. At the lowest investigated shear
rate, $0.01\;s^{-1}$, the velocity profile is linear (see
Fig.\ref{fig:flowvelocitycurve}b). Increasing the shear rate to
$0.1 \; s^{-1}$, inside the stress plateau region, a banded
structure can be recognized and the velocity profile shows a
characteristic kink, as can be seen from
Fig.\ref{fig:flowvelocitycurve}b. In the investigated overall
shear-rate range, the average shear rate in the high shear-rate
band is twice that of the lower shear-rate band. The
fraction of the gap occupied by the high shear-rate band increases
with the overall shear rate, and for shear rates higher than $1\;
s^{-1}$ the low shear-rate band disappears and a linear profile is
re-established.

\begin{figure}\centering
\includegraphics[width=1\textwidth]{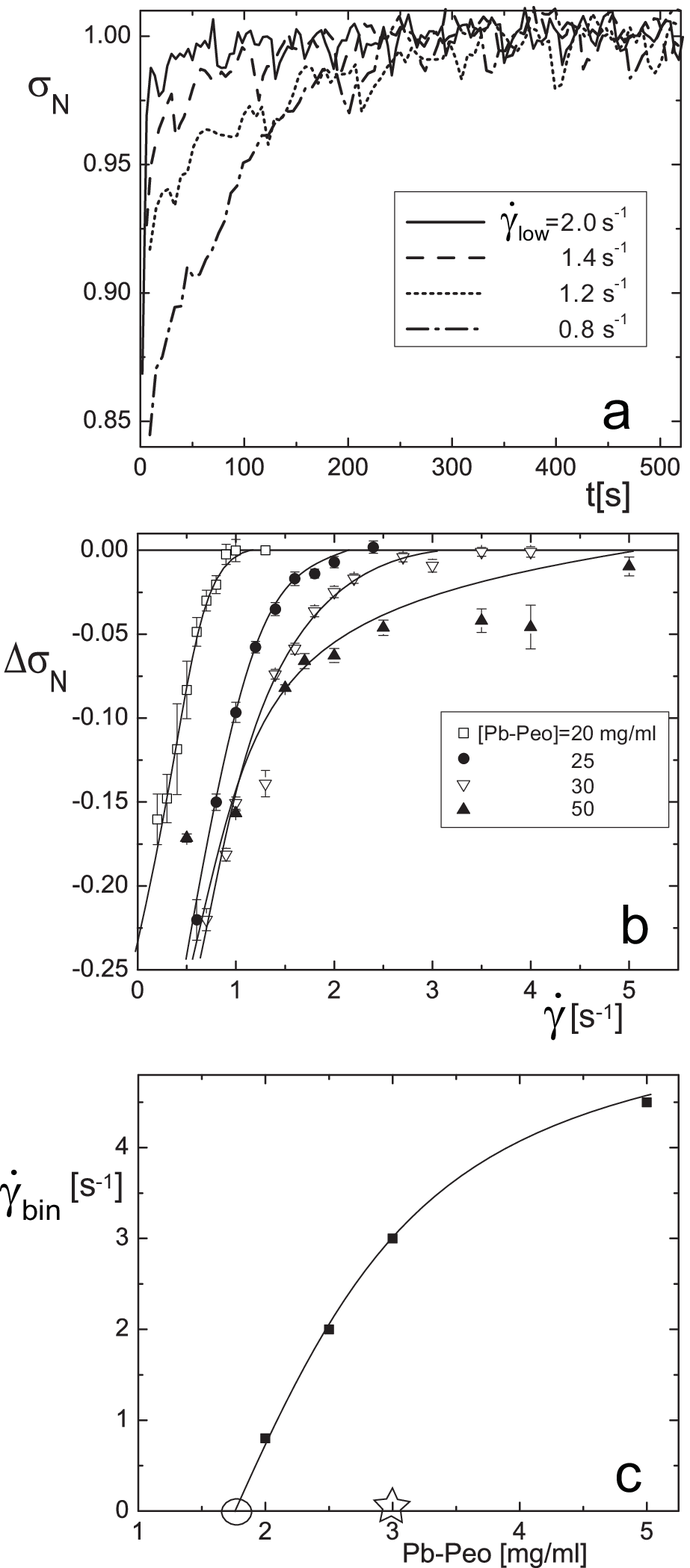}
\caption{(a) The response of the normalized stress
$\sigma_N(t)=\sigma(t)/\sigma(t\rightarrow\infty)$ to shear rate
quenches from the fully nematic state into the biphasic region.
The initial shear rate was $\gdot = 7 \; s^{-1}$ and the low shear
rates were varied between $\gdot = 0.8\; s^{-1}$ (bottom) and
$\gdot = 2.0\; s^{-1}$ (top). (b) The magnitude of the stress
response $\Delta\sigma_N$, obtained from the fit to
$\sigma_N(t)=1-\Delta\sigma_Ne^{-t/\tau}$ as a function of the
shear rate. Lines are guide to the eye. (c) The resulting binodal points obtained from the shear
rate in Fig.b where $\Delta\sigma_N$ becomes zero. The circle
indicates the equilibrium I-N binodal, that is, the binodal point
in the absence of flow. The line is a guide to the eye representing the non-equilibrium binodal. The open star indicates the location of
the spinodal at zero shear rate.} \label{fig:quenchdown}
\end{figure}

In order to locate the isotropic-nematic binodal, i.e. the point
where the isotropic phase becomes meta-stable, rheology is a very
useful tool as the viscosity of the micellar solution is very
sensitive to the local orientation of the worms. To exploit the
large difference between the viscosity of the isotropic and
nematic phase, we performed step-down experiments in the
concentration region between $2\; \%$ and $5\; \%$. As we have
shown in an earlier paper on rod-like viruses \cite{Lettinga04a},
the viscosity of the system will increase in time when the system
is quenched from a high shear-rate, where the nematic phase is
stable, to a lower shear rate where the nematic phase becomes
meta- or unstable. Fig.\ref{fig:quenchdown}a shows an example of
the normalized stress
$\sigma_N(t)=\sigma(t)/\sigma(t\rightarrow\infty)$ (where $\sigma$
is the shear stress) as a function of time after the shear rate
was quenched from $7\; s^{-1}$ to a final value ranging from $2$
to $0.8\; s^{-1}$. The curves are fitted to a single exponential
$\sigma_N(t)=1-\Delta\sigma_Ne^{-t/\tau}$, where $\Delta\sigma_N$
depends on the fraction of the formed isotropic phase, which tends
to zero at the binodal point. Thus, for each concentration the
binodal point was determined as the shear rate at which
$\Delta\sigma_N$ vanishes (see Fig.\ref{fig:quenchdown}b). The
resulting binodal points are plotted in Fig.\ref{fig:quenchdown}c.
This figure constitutes the low concentration branch of the
non-equilibrium binodal for the Pb-Peo block copolymer system. The
equilibrium I-N binodal, in the absence of flow, is found to be
located at $[Pb-Peo]=1.7 \;\pm 0.1\; \%$. The open star in Fig.\ref{fig:quenchdown} indicates the location
of the spinodal at zero shear rate. How this spinodal point was
determined will be discussed later.

As we are dealing with a
system that can also be described as flexible rods, we know from
e.g. Chen \cite{chen93} that the I-N phase coexistence region is
very broad and thus that the location of the I-N spinodal can be
found at a significantly higher concentration than the I-N
binodal. In case of rigid rods, the collective rotational
diffusion becomes very small on approach of the spinodal point, as
discussed in section \ref{Theory}. For semi-flexible chains, the
rotational motion of the Kuhn-segments will become very slow on
approach of the spinodal. We will now employ dynamic experiments
to access this slowing down. We want to do this not only on the
macroscopic level, i.e. by rheology, but also on the microscopic
level, in order to establish a link between the behavior of
Kuhn-segments and the measured stress in the system.

\begin{figure}\center
\includegraphics[width=1\textwidth]{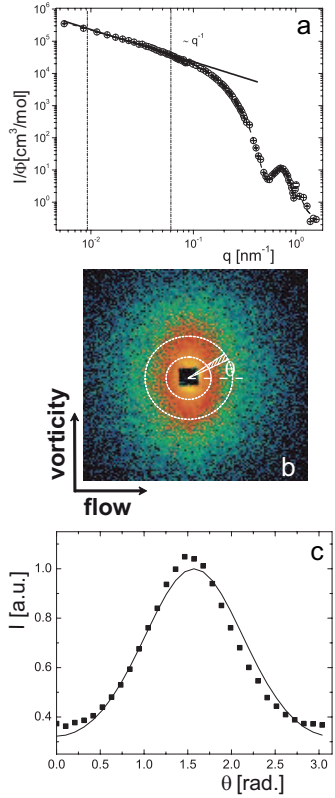}
\caption{(a) Angle averaged SANS curve at zero shear. The full
line indicates the $q$-range where a $q^{-1}$-dependence is found,
typical for rods. (b) Scattering pattern of $1\; \%$ Pb-Peo in
deuterated water at shear rate $\gdot=1\;s^{-1}$ in the
flow-velocity plane. The dashed lines indicate the $q$-range that
is used to obtain the azimuthal intensity profile as plotted in
(c). Here $\theta$ is the angle with the shear flow,  and the full
line indicates a fit to Eq.\ref{Eq_Iq}.} \label{fig_IvsQ}
\end{figure}

\subsection{SANS on quiescent and stationary sheared samples}

The quiescent dispersion of Pb-Peo micelles has an angle-averaged
scattering pattern as plotted in Fig.\ref{fig_IvsQ}a. At low $Q$
values, the scattering curve shows a $I\sim q^{-1}$ dependence,
typical for rods. The transition from $I\sim q^{-1}$ to a $I\sim
q^{-2}$ dependence that is expected for worm-like micelles is
outside the experimental window. This shows that the persistence
length of the worms is at least $500\;nm$, in agreement with
previous experiments on the same system\cite{Won99}. For this
reason the data could be fitted with the form factor of a long
cylinder. The details of the fitting procedure are beyond the
scope of this paper, and will be described elsewhere. The main
point is that the cylinders are assumed to have a uniform core and
a shell with an exponential density profile, i.e. density$\sim
(1-r)e^{(-\alpha r)}$, where
$r=\frac{r-\sigma_{core}}{\sigma_{shell}}$. From the fitting, the
core and shell radii $\sigma_{core}$ and $\sigma_{shell}$, the
aggregation number per unit length and the exponent $\alpha$ have
been obtained. The numerical values of these parameters are given
in Table \ref{table:fit}. The values for the cylinder cross
section is in agreement with that already reported in
literature\cite{Won99}.

\begin{table}[!hptb]
\centering
\begin{tabular}{ccccc}
$N_{ag}/nm^{-1}$ & $\sigma_{core}$/nm & $\sigma_{shell}$/ nm & $\alpha$ \\
\hline
\hline

26.5 & 6.4 & 7.8 & 3.65  \\
\hline
\end{tabular}
\caption{Structural micelle characteristics as obtained from
fitting of the SANS curve in Fig.\ref{fig_IvsQ}}
\label{table:fit}
\end{table}

Fig.\ref{fig_IvsQ}b shows a typical scattering pattern of Pb-Peo
under shear conditions (with $\gdot=1\;s^{-1}$), which shows the
shear-induced anisotropic structure. This can be more clearly seen
in the azimuthal intensity profile, as plotted in
Fig.\ref{fig_IvsQ}c, which is obtained from the part of the
scattering pattern in Fig.\ref{fig_IvsQ}b where the scattered
intensity is proportional to $q^{-1}$ (the area in between the to
circles in Fig\ref{fig_IvsQ}b). Assuming a Maier-Saupe type of
orientation distribution function, the azimuthal scattered
intensity $I(Q,\theta)$ from the nematic phase is generally well
described by\cite{Picken90},
\begin{equation}\label{Eq_Iq}
I(Q,\theta)\;\sim\;\exp\left\{ \beta P_2(\theta)-1\right\}\;,
\end{equation}
where the parameter $\beta$ describes the width of the intensity
profile and $P_{2}$ is the second order Legendre polynomial.
The solid line in Fig.\ref{fig_IvsQ}c shows an example of a fit of the expression in
eq.(\ref{Eq_Iq}) with the experimental data as obtained from the scattering pattern in Fig.\ref{fig_IvsQ}b.
The scalar order parameter $\langle P_{2}(\theta)\rangle$ can then be
calculated from,
\begin{equation}\label{Eq_P2exp}
\langle P_{2}(\theta)\rangle=\frac{\int_0^{\pi}\exp\left\{ \beta P_2(\theta)\right\} \;P_2(\theta)\sin(\theta)d\theta}
{\int_0^{\pi}\exp\left\{ \beta P_2(\theta)\right\}\;\sin(\theta)d\theta}\;.
\end{equation}
In this way, the order parameter $\langle P_{2}(\theta)\rangle$
can be obtained from scattering data for each shear rate at different concentrations. As
expected, flow-induced orientation of the cylindrical micelles is observed.

\begin{figure}\center
\includegraphics[width=.5\textwidth]{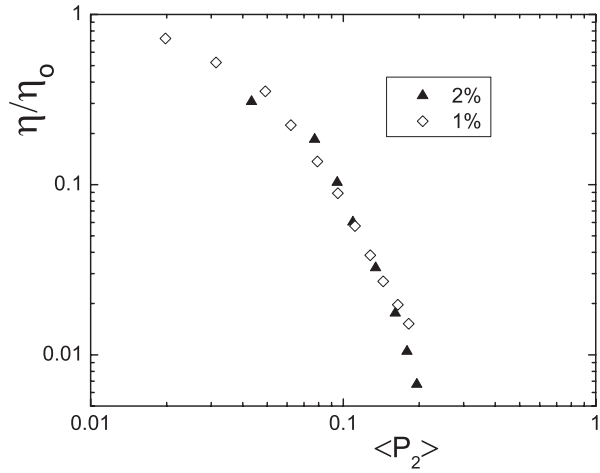}
\caption{The scaled viscosity $\eta$ (with $\eta_{0}$ the viscosity at zero shear rate)
versus the orientational order parameter $\langle
P_{2}\rangle$.}\label{fig_P2vsSR}
\end{figure}

In ref.\cite{Foerster05} it is suggested that the shear viscosity
is a universal function when plotted against the orientational
order parameter, independent of concentration. We indeed find such
a behaviour for our Pb-Peo system, as can be seen from
Fig.\ref{fig_P2vsSR}. For the two concentrations of $1$ and
$2\;\%$, the two curves fall on top of each other. Contrary to
ref.\cite{Foerster05}, we do not find a linear dependence of the
viscosity on the order parameter, probably due to the fact that we
also used data at shear rates lower than those where the stress
plateau occurs.

\begin{figure}\center
\includegraphics[width=1\textwidth]{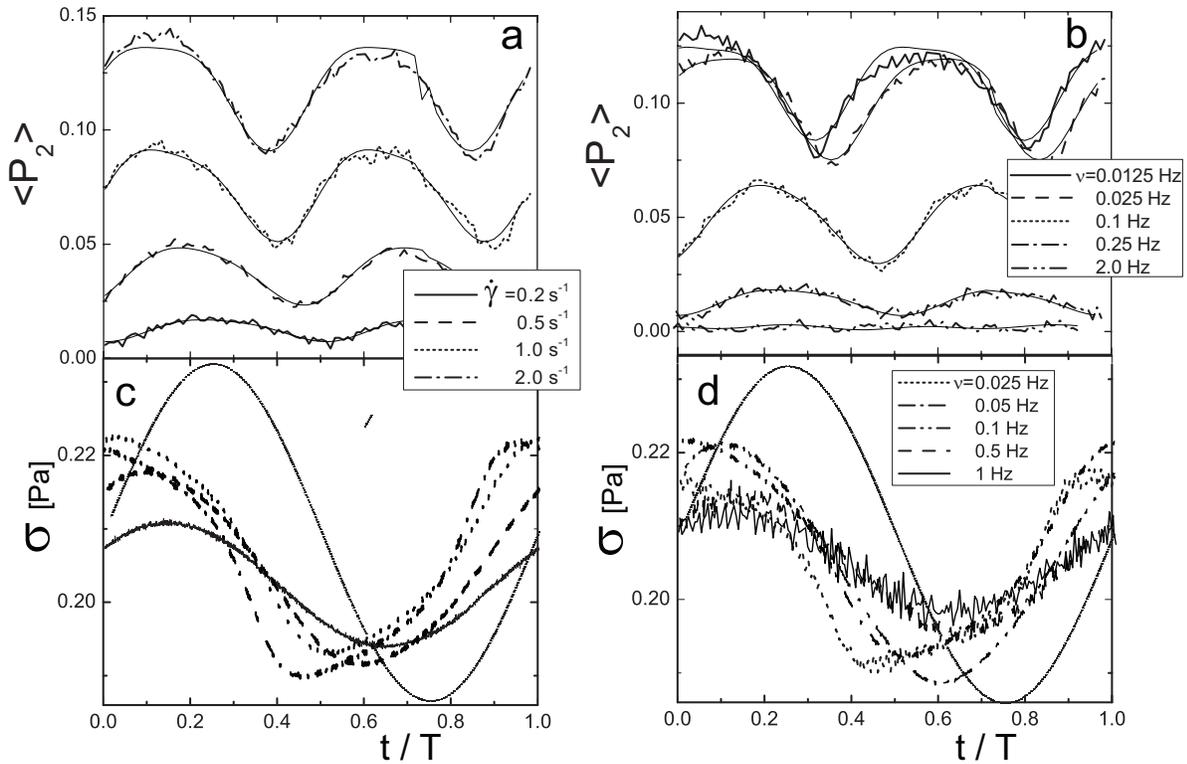}
\caption{Time-dependent response of the orientational order
parameter $\langle P_{2}\rangle$ (a,b) and the stress (c,d) to an
oscillatory shear flow at a shear rate of
$\gdot_{max}=1.0\;s^{-1}$ (a,c) and a frequency of $0.05\;Hz$
(b,d) at a concentration of $2\; \%$ Pb-Peo. The thin dotted
curves indicate the applied shear rate. The time $t$ is scaled
with the period $T$ of oscillation.} \label{fig:responseExp}
\end{figure}

\begin{figure}\center
\includegraphics[width=1\textwidth]{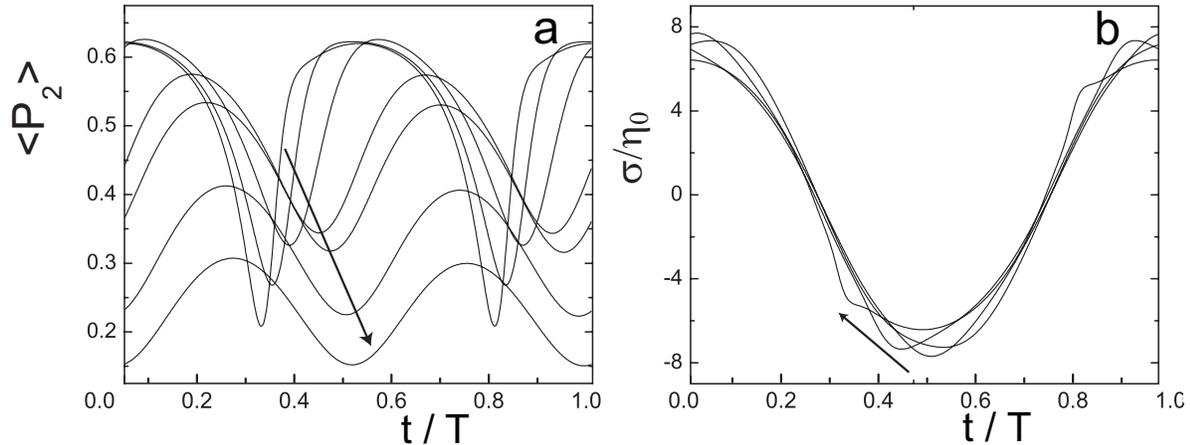}
\caption{Theoretical predictions for the response of (a) the
stress and (b) the order parameter $\langle P_{2}\rangle$ for
$\Omega$ varying from $3$ to $60$. The arrows indicates increasing
$\Omega$. The effective Peclet number is $Pe_{eff}=75$ and the
concentration is $\frac{L}{d}\varphi =3.3$. The time $t$ is scaled
with the period $T$ of oscillation.} \label{fig:responseTheo}
\end{figure}

\begin{figure}\centering
\includegraphics[width=1\textwidth]{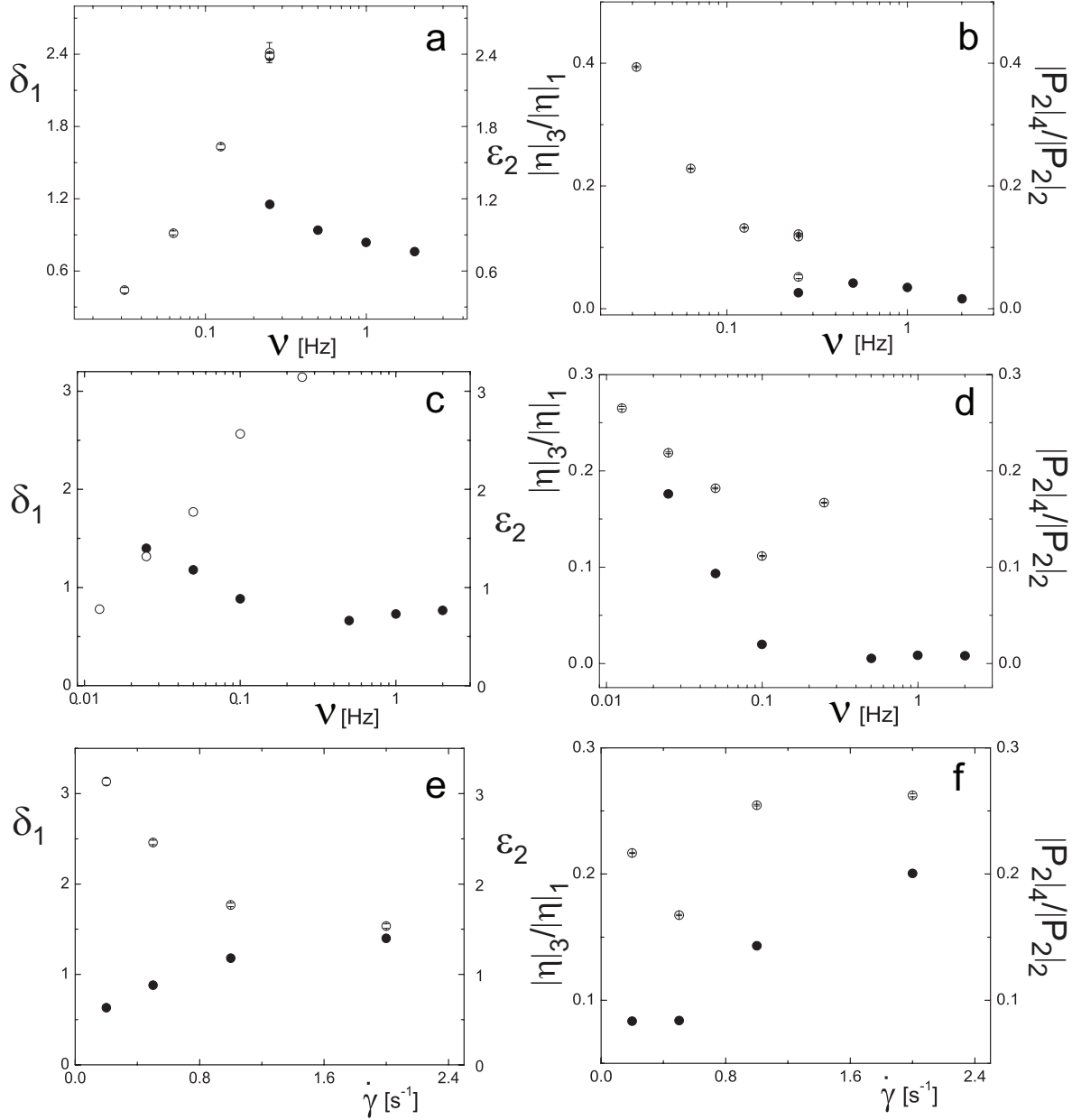}
\caption{Results of the Fourier analysis using
eq.\ref{Eq_FourierP2} for $<P_2>(t)$ (open symbols) and
eq.\ref{Eq_FourierSig} for the stress (filled symbols) at $1\; \%$
(a,b) and $2\; \%$ (c-f) Pb-Peo. (a,c,e) are plots of the phase
shift $\delta_1$ for the stress and  $\epsilon_2$ for $\langle
P_{2}\rangle$, while (b,d,f) are plots of $|\eta|_3/|\eta|_1$ for
the stress and $|P_{2}|_4/|P_{2}|_2$ for $\langle P_{2}\rangle$),
which quantities measure the departure from linear response. In
(a-d), the shear rate is fixed to $\gdot_{max}=4.0\;s^{-1}$ for
$1\;\%$ Pb-Peo and $\gdot_{max}=1.0\;s^{-1}$ for $2\;\%$ Pb-Peo.
In (e,f) the frequency is fixed to $\nu = 0.05\; Hz$.}
\label{fig:Fourier}
\end{figure}

\subsection{Dynamic experiments}

Oscillatory shear rate experiments were performed for
concentrations lower and around the I-N equilibrium binodal point,
i.e. between $0.5$ and $2\; \%$. As for steady-state measurements,
the order parameter $\langle P_{2}\rangle$ can be calculated from
SANS experiments according to eq.\ref{Eq_Iq} and \ref{Eq_P2exp} at
each time during an oscillation. In this way we probe the
time dependence of the orientational order parameter $\langle
P_{2}\rangle$. In order to compare and relate the orientational
response with the change in the stress, the stress response was
also recorded and analyzed by Fast Fourier Transform Rheology
experiments on samples in a somewhat broader concentration range
between $0.5$ and $2.5\; \%$.

In Fig.\ref{fig:responseExp} we plot the time-dependent response
of $\langle P_{2}\rangle$ (a,b) and the stress (c,d) of a $2\;\%$
sample. In Figs.\ref{fig:responseExp}a,c, the response for
different shear-amplitudes $\dot{\gamma}_{0}$ (see
eq.\ref{shearrate})) is shown, where the maximum shear rate during
an oscillation is kept constant by adjusting the frequency. In
Figs.\ref{fig:responseExp}b,d, the response for different
frequencies is shown, where again the maximum shear rate during an
oscillation is kept constant, but now by adjusting the
shear-amplitude. The first thing to note is that the order
parameter oscillates with twice the frequency of the applied shear
rate, even for low shear rates where the stress response is linear
in the shear rate. The reason for this is that the scattering
experiments probe the flow-vorticity plane, so that the measured
order parameter characterizes the orientational order within that
plane. As already discussed in section \ref{Theory}, there is no
linear response of the order parameter in this plane, and the
leading response is quadratic in the shear rate. This results in
the double-frequency response of the probed projection of
orientational order. The experimental trends are in good
qualitative agreement with the theoretical calculations based on
eqs.(\ref{Eq_stress},\ref{eq_St}) as can be seen from
Fig.\ref{fig:responseTheo}, where Fig.\ref{fig:responseTheo}a
should be compared to the experimental results in
Fig.\ref{fig:responseExp}b, and Fig.\ref{fig:responseTheo}b to
Fig.\ref{fig:responseExp}d. The theoretical curves have the same
form as the experimental curves, exhibiting similar trends on
changing frequency and shear-amplitude. In order to quantify the
dynamic response we analyze this response on the basis of the
Fourier modes as given in
eqs.(\ref{Eq_FourierSig},\ref{Eq_FourierP2}) for the stress and
(flow-vorticity projected) orientational order parameter $\langle
P_{2}\rangle$, respectively. The experimental phase shifts for the $2\;\%$ sample are shown in
Fig.\ref{fig:Fourier}a,c and e. The Fourier amplitude ratios that
measure the departure from linear response, $|P_{2}|_4/|P_{2}|_2$
for $\langle P_{2}\rangle$ and $|\eta|_3/|\eta|_1$ for the stress,
are plotted in Fig.\ref{fig:Fourier}b,d and f.

\begin{figure}\center
\includegraphics[width=.5\textwidth]{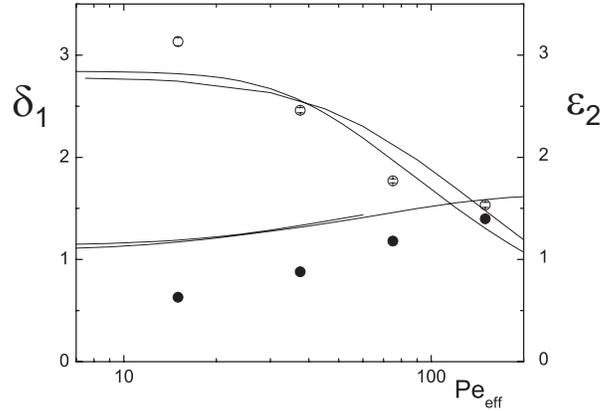}
\caption{Shear rate dependence of the  phase shifts $\delta_1$ and
$\epsilon_2$ for the theoretically calculated response of the stress
(bottom curves) and the orientational order parameter $\langle
P_{2}\rangle$ (top curves), respectively, at a scaled volume
fraction of $\frac{L}{d}\phi=10/3$ (solid line) and
$\frac{L}{d}\varphi=5/3$ (dashed line). The effective Deborah
number is $\Omega_{eff}=24$. The symbols give the experimental
response for the stress (solid) and $\langle P_{2}\rangle$ (open)
at $2\; \%$ Pb-Peo, scaled with the orientational diffusion
coefficient at infinite dilution with a value of
$D_r=0.04\;s^{-1}$ and $C=3$.}
\label{fig:FourierTheory}
\end{figure}

\begin{figure}\center
\includegraphics[width=1.\textwidth]{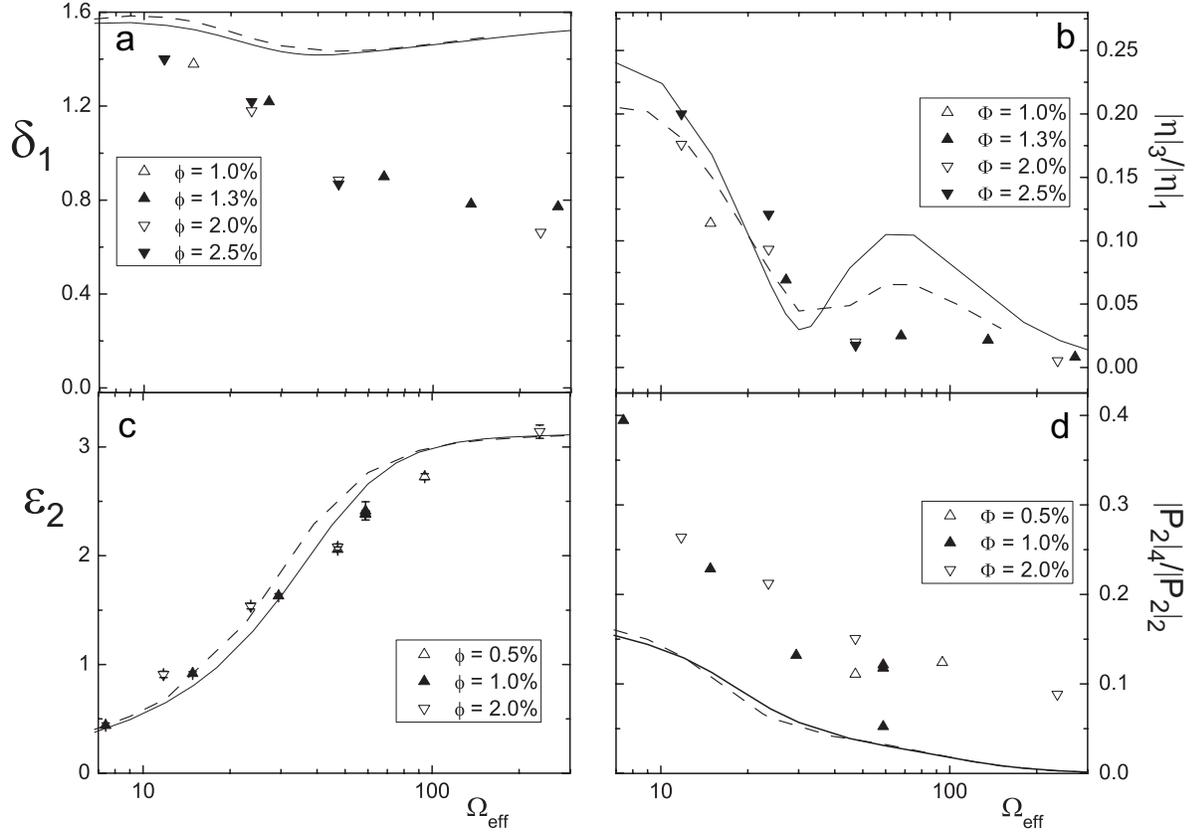}
\caption{ The Phase shifts (a,c) and non-linearity (b,d) for the
stress (a,b) and $\langle P_{2}\rangle$ (c,d) versus the Deborah
number. The symbols indicate the experiments for different
concentrations. The solid lines give the theoretical responses for
$\frac{L}{d}\varphi=10/3$ and the dashed line for
$\frac{L}{d}\varphi=5/3$. For the scaling of the experimental
frequency we used a value for the orientational diffusion
coefficient at infinite dilution of $D_r=0.04\;s^{-1}$ and $C=3$
was used, see Eq. \ref{Eq_scaleD}. $Pe_{eff}\;=\;250$ for all
data.}\label{fig:Scaling}
\end{figure}

As mentioned in the theory section \ref{Theory}, the rate at which
a dispersion of rods relaxes close to the spinodal point is
determined by effective diffusion coefficient $D_{eff}$ given by
Eq.\ref{Eq_deff}. There are two unknown parameters in this
equation, namely the spinodal concentration, i.e. the
concentration where $\frac{L}{d}\phi\;=\;4$, and the rotational
diffusion at infinite dilution $D_r$. When critical slowing down
is at the origin of the difference in dynamic response for various
concentrations, we should find scaling when repsonse functions are
plotted against effective quantities, like the effective Peclet
number in eq.(\ref{Eq_Peclet}) and the effective Deborah number in
eq.(\ref{Eq_Deborah}). In order to test such a scaling for
relatively low shear rates, we need to know the concentration
where the I-N spinodal in the absence of flow is located. In view
of our expression (\ref{Eq_deff}) for the effective rotational
diffusion coefficient, we will use the following similar form for
the effective diffusion coefficient of the Pb-Peo system,
\begin{equation}\label{Eq_scaleD}
D_r^{eff}=D_r\left\{1-[Pb-Peo]/C\right\}\;,
\end{equation}
where, as before $[Pb-Peo]$ is the concentration of Pb-Peo, and
$C$ is a scaling parameter that determines the location of the I-N
spinodal. For a given value of the parameter $C$, the effective
Peclet and Deborah numbers are calculated from
eq.(\ref{Eq_Peclet}) and eq.(\ref{Eq_Deborah}).

As can be seen from Fig.\ref{fig:Scaling}, all experimental data
for phase shifts and non-linear response functions collapse on a
single curve when $C$ is taken equal to $3$. This is true for both
the stress response as well as for the response of the
orientational order parameter $\langle P_{2}\rangle$ (projected on
the flow-vorticity plane). This is in accord with the idea that
the concentration dependence of the response of both orientational
order as well as the stress is related to critical slowing down.
Thus, in terms of polymer concentration, the spinodal point is
located at $[Pb-Peo]_{spin}=3\;\%$. The spinodal point is
indicated by the open star in Fig.\ref{fig:quenchdown}c. This
spinodal concentration seems to be in accord with the somewhat
lower binodal concentration of $1.7\;\%$ in the absence of flow.

The spinodal concentration can be estimated from the length,
thickness and volume fraction of the wormlike micelles, neglecting
the effect of the flexibility. As discussed before, the structural
parameters of the worm-like micelles have been derived from
fitting of scattering data (see Table \ref{table:fit}). In this
way we can estimate the location of the I-N spinodal by equating
both expressions (\ref{Eq_deff}) and (\ref{Eq_scaleD}) for the
effective diffusion coefficient. Since $\phi= 2.73\times
[Pb-Peo]$, $d=14.215.6\; nm$ and $L=1\; \mu m$ ($L$ obtained from DLS and
microscopy data, to be published ) we find that,
\begin{equation}\label{Eq_scaleD1}
[Pb-Peo]_{spin}\;=\;\frac{4}{2.73}\times \frac{d}{L}=2.1\; \%\;,
\end{equation}
which, in view of the neglect of flexibility in obtaining this
number, is in reasonable agreement with the value obtained from
the dynamic experiments.

In order to compare the experimental results with theory, the
rotational diffusion at infinite dilution $D_r$ needs to be
determined. To do so we determine the Deborah number
$\Omega_{eff}$ for which the limiting values for the phase shifts
for $Pe_{eff}\rightarrow 0$ as found in the
experiments is reproduced. As can be seen in
Fig.\ref{fig:FourierTheory} there is a reasonable comparison,
especially for $\epsilon_{2}(Pe_{eff}\rightarrow 0)$, between the
theoretical calculation using $\Omega_{eff}\;=\;24$ and the
experimental frequency of $0.05\;Hz$ at a concentration of $2\;
\%$. Since we know from the scaling that $C=3$, it follows that
$\frac{L}{d}\varphi=10/3$ for this concentration, and thus, with
Eq.\ref{Eq_Deborah} we find that $D_r\;=\;0.04\;s^{-1}$. This
number, together with the dimensionless concentration
$\frac{L}{d}\varphi$, was used in the scaling of the frequency and
shear rates in Fig.\ref{fig:Scaling}. In this figure, as well as
in Fig. \ref{fig:FourierTheory}, the theoretical validity of the
scaling argument is tested by calculating the dynamic response at
two different dimensionless concentrations
$\frac{L}{d}\varphi=10/3$ and $5/3$, having the same distance to
the spinodal point as the $2\; \%$ and $1\; \%$ samples.

The theoretical frequency dependence of phase shifts and
non-linear response functions exhibit the same features as the
experimental results, as can be seen from Fig.\ref{fig:Scaling} :
the functional form of both is reproduced and the absolute values
are in qualitative agreement. Due to the neglect of flexibility, a
quantitative agreement is not expected. What is more important, however,
is that the functional variation with the effective frequency is
the same for both experiment and theory. We can therefore draw the
important conclusion that the scaling with eq.(\ref{Eq_scaleD}) in
the experiments is justified. In other words, the flow response of
the Pb-Peo system scales with the distance from the spinodal
point.

The correspondence between theory and experiment is especially
satisfactory for the frequency dependence of the phase shift in
$\langle P_{2}\rangle$, $\epsilon_2$ in Fig.\ref{fig:Scaling}c, and
the non linearity in the stress, given by  $|\eta|_3/|\eta|_1$ in
Fig.\ref{fig:Scaling}b. This correspondence confirms the choice of
$D_r=0.04\;s^{-1}$. The experimental phase shift $\delta_1$ in the
stress, given in Fig.\ref{fig:Scaling}a, however, shows a more
pronounced frequency dependence as predicted by theory. Concerning
the phase shifts, it is interesting to note that at low
frequencies, $\langle P_{2}\rangle$ is in phase with the applied
shear field $\gdot\propto \frac{d\gamma}{dt}\propto cos(\omega t)$
and $\epsilon_2=0$, while at high frequencies $\epsilon_2\rightarrow
\pi$. For the stress we observe that at low frequencies
$\delta_1\;=\;\pi/2$, corresponding to fluid-like behavior, while
$\delta_1$ decreases with increasing frequency, but never reaches
$0$, which value corresponds to solid-like behavior. The variation
of $\epsilon_2$ is twice that of $\delta_1$ due to the fact that
$\langle P_{2}\rangle(t)\sim\gdot^2$, as discussed before.

The frequency dependence of the non-linear response functions show
that with increasing frequency the system becomes more linear (at
the cost of an increasing phase shift). The linear-response regime
therefore extends up to larger shear rates when the frequency
increases. The reason for this is that at high frequencies,
microstructural order is not able any more to fully respond to the
external field.

There is a considerable discrepancy between the value of the
orientational diffusion coefficient $D_r$ at infinite dilution
that one would calculate for the length and thickness of the worms
from well-known expressions for stiff rods \cite{Tirado80} and the
value found in our experiments. It is unclear whether this is the
result of the flexibility of the rods. Another source for this
discrepancy might be that the theory neglects dynamical
correlations. In the derivation of
eqs.(\ref{eq_St},\ref{Eq_stress}), the rod-rod pair-correlation
function is taken equal to the Boltzmann exponential of the
pair-interaction potential. This is asymptotically exact for very
long and thin hard rods for the calculation of thermodynamic
quantities of rod suspensions. For dynamical processes (with or
without shear flow), however, such an approximation for the
pair-correlation is approximate, and particularly neglects
dynamical correlations. Simulations have shown that such
correlations are of importance, at least for fast dynamical
processes\cite{Tao06c}. The simulations show that critical slowing
down is enhanced by dynamical correlations. This might explain the
above mentioned discrepancy between theory and experiment. This is
a subject for future investigations.

\section{Conclusion}
The aim of this paper is to find the microscopic mechanism of
the strong shear-thinning behavior of giant wormlike micelles consisting
of Pb-Peo block copolymers. The dynamics of the stress is probed by
dynamic shear experiments in the linear and non-linear regime
using Fourier Transfer Rheology. The dynamics of the orientational order parameter under oscillatory flow
is studied with a newly developed time-resolved neutron scattering set up.
It is shown that critical slowing down of orientational Brownian motion due to the vicinity of the isotropic-nematic spinodal is responsible for the shear thinning behaviour. The response functions for different concentrations are indeed identical when plotted against an effective Deborah number that accounts for critical slowing down. In a certain shear-rate range, shear thinning is so strong that gradient shear banding occurs, where flow profiles have been measured with heterodyne light scattering. The location of the binodal in the shear rate versus concentration plane is determined by step-down rheology, and the spinodal concentration in the absence of flow is obtained from the scaling behaviour of response functions. Both the measured linear and non-linear stress response and order parameter response are in qualitative agreement with a theory for stiff rods that includes critical slowing down on approach of the isotropic-nematic spinodal. The comparison with theory, however, is qualitative since the theory neglects flexibility. Another possible reason for deviations between theory and experiments might be that the theory neglects dynamical correlations, which have been shown by simulations to enhance critical slowing down. In surfactant worm-like micellar systems, shear thinning can also be due to breaking and stress-induced merging of worms. The breaking and merging of worms can give rise to strong shear thinning by itself, and can give rise to shear banding in the absence of critical slowing down, far away from the spinodal. Scission and stress-induced merging do probably not play a role in the Pb-Peo block copolymer system that we studied here.

\textbf{Acknowledgement} We acknowledge the Transregio
Sonderforschungsbereich TR6018 "Physics of Colloidal Dispersions in External Fields" for financial support. We are grateful
to J\"{o}rg Stellbrink, Peter Lang and Aggeliki Tsigkri for the
information on the particle characteristics.


\section*{References}
\begin{thebibliography}{22}
\expandafter\ifx\csname natexlab\endcsname\relax\def\natexlab#1{#1}\fi
\expandafter\ifx\csname bibnamefont\endcsname\relax
  \def\bibnamefont#1{#1}\fi
\expandafter\ifx\csname bibfnamefont\endcsname\relax
  \def\bibfnamefont#1{#1}\fi
\expandafter\ifx\csname citenamefont\endcsname\relax
  \def\citenamefont#1{#1}\fi
\expandafter\ifx\csname url\endcsname\relax
  \def\url#1{\texttt{#1}}\fi
\expandafter\ifx\csname urlprefix\endcsname\relax\def\urlprefix{URL }\fi
\providecommand{\bibinfo}[2]{#2}
\providecommand{\eprint}[2][]{\url{#2}}

\bibitem[{\citenamefont{Berret}(2004)}]{Berret04}
\bibinfo{author}{\bibfnamefont{J.-F.} \bibnamefont{Berret}},
  \emph{\bibinfo{title}{Molecular Gels}} (\bibinfo{publisher}{Elsevier},
  \bibinfo{year}{2004}), chap. \bibinfo{chapter}{Rheophysics of Wormlike
  Micelles}.

\bibitem[{\citenamefont{Manneville}(2008)}]{Manneville08}
\bibinfo{author}{\bibfnamefont{S.}~\bibnamefont{Manneville}},
  \bibinfo{journal}{Rheol. Acta} \textbf{\bibinfo{volume}{47}},
  \bibinfo{pages}{301318} (\bibinfo{year}{2008}).

\bibitem[{\citenamefont{Cates and Candau}(1990)}]{Cates90a}
\bibinfo{author}{\bibfnamefont{M.~E.} \bibnamefont{Cates}} \bibnamefont{and}
  \bibinfo{author}{\bibfnamefont{S.~J.} \bibnamefont{Candau}},
  \bibinfo{journal}{J. Phys.: Condens. Matter} \textbf{\bibinfo{volume}{2}},
  \bibinfo{pages}{6869} (\bibinfo{year}{1990}).

\bibitem[{\citenamefont{Briels et~al.}(2004)\citenamefont{Briels, Mulder, and
  den Otter}}]{Briels04}
\bibinfo{author}{\bibfnamefont{W.~J.} \bibnamefont{Briels}},
  \bibinfo{author}{\bibfnamefont{P.}~\bibnamefont{Mulder}}, \bibnamefont{and}
  \bibinfo{author}{\bibfnamefont{W.~K.} \bibnamefont{den Otter}},
  \bibinfo{journal}{J. Phys.: Condens. Matter 16 (2004) S3965S3974}
  \textbf{\bibinfo{volume}{16}}, \bibinfo{pages}{S3965S3974}
  (\bibinfo{year}{2004}).

\bibitem[{\citenamefont{Won et~al.}(1999)\citenamefont{Won, Davis, and
  Bates}}]{Won99}
\bibinfo{author}{\bibfnamefont{Y.-Y.} \bibnamefont{Won}},
  \bibinfo{author}{\bibfnamefont{H.~T.} \bibnamefont{Davis}}, \bibnamefont{and}
  \bibinfo{author}{\bibfnamefont{F.~S.} \bibnamefont{Bates}},
  \bibinfo{journal}{Science} \textbf{\bibinfo{volume}{283}},
  \bibinfo{pages}{960} (\bibinfo{year}{1999}).

\bibitem[{\citenamefont{Lund et~al.}(2006)\citenamefont{Lund}}]{Lund06}
\bibinfo{author}{\bibfnamefont{R.} \bibnamefont{Lund}},
  \bibinfo{author}{\bibfnamefont{L.}~\bibnamefont{Willner}}
  \bibinfo{author}{\bibfnamefont{D.}~\bibnamefont{Richter}}
  , \bibnamefont{and}
  \bibinfo{author}{\bibfnamefont{E. E.} \bibnamefont{Dormidontova}},
  \bibinfo{journal}{Macromolecules} \textbf{\bibinfo{volume}{39}},
  \bibinfo{pages}{4566} (\bibinfo{year}{2006}).

\bibitem[{\citenamefont{Denkova et~al.}(2008)\citenamefont{Denkova, Mendes, and
  Coppens}}]{Denkova08}
\bibinfo{author}{\bibfnamefont{A.~G.} \bibnamefont{Denkova}},
  \bibinfo{author}{\bibfnamefont{E.}~\bibnamefont{Mendes}}, \bibnamefont{and}
  \bibinfo{author}{\bibfnamefont{M.-O.} \bibnamefont{Coppens}},
  \bibinfo{journal}{J. Phys. Chem. B} \textbf{\bibinfo{volume}{112}},
  \bibinfo{pages}{793} (\bibinfo{year}{2008}).

\bibitem[{\citenamefont{{F\"o}rster et~al.}(2005)\citenamefont{{F\"o}rster,
  Konrad, and Lindner}}]{Foerster05}
\bibinfo{author}{\bibfnamefont{S.}~\bibnamefont{{F\"o}rster}},
  \bibinfo{author}{\bibfnamefont{M.}~\bibnamefont{Konrad}}, \bibnamefont{and}
  \bibinfo{author}{\bibfnamefont{P.}~\bibnamefont{Lindner}},
  \bibinfo{journal}{Phys. Rev. Lett.} \textbf{\bibinfo{volume}{94}},
  \bibinfo{pages}{017803} (\bibinfo{year}{2005}).

\bibitem[{\citenamefont{Dhont and Briels}(2003{\natexlab{a}})}]{Dhont03c}
\bibinfo{author}{\bibfnamefont{J.~K.~G.} \bibnamefont{Dhont}} \bibnamefont{and}
  \bibinfo{author}{\bibfnamefont{W.~J.} \bibnamefont{Briels}},
  \bibinfo{journal}{Colloid Surface A} \textbf{\bibinfo{volume}{213}},
  \bibinfo{pages}{131} (\bibinfo{year}{2003}{\natexlab{a}}).

\bibitem[{\citenamefont{Doi and Edwards}(1986)}]{Doi86}
\bibinfo{author}{\bibfnamefont{M.}~\bibnamefont{Doi}} \bibnamefont{and}
  \bibinfo{author}{\bibfnamefont{S.~F.} \bibnamefont{Edwards}},
  \emph{\bibinfo{title}{The Theory of Polymer Dynamics}}
  (\bibinfo{publisher}{Clarendon Press, Oxford}, \bibinfo{year}{1986}).

\bibitem[{\citenamefont{Dhont and Briels}(2003{\natexlab{b}})}]{Dhont03b}
\bibinfo{author}{\bibfnamefont{J.~K.~G.} \bibnamefont{Dhont}} \bibnamefont{and}
  \bibinfo{author}{\bibfnamefont{W.~J.} \bibnamefont{Briels}},
  \bibinfo{journal}{J. Chem. Phys.} \textbf{\bibinfo{volume}{118}},
  \bibinfo{pages}{1466} (\bibinfo{year}{2003}{\natexlab{b}}).

\bibitem[{\citenamefont{Allgaier et~al.}(1997)\citenamefont{Allgaier, Poppe,
  Willner, and Richter}}]{Allgaier97}
\bibinfo{author}{\bibfnamefont{J.}~\bibnamefont{Allgaier}},
  \bibinfo{author}{\bibfnamefont{A.}~\bibnamefont{Poppe}},
  \bibinfo{author}{\bibfnamefont{L.}~\bibnamefont{Willner}}, \bibnamefont{and}
  \bibinfo{author}{\bibfnamefont{D.}~\bibnamefont{Richter}},
  \bibinfo{journal}{Macromolecules} \textbf{\bibinfo{volume}{30}},
  \bibinfo{pages}{1582} (\bibinfo{year}{1997}).

\bibitem[{\citenamefont{Kohlbrecher and Wagner}(2000)}]{Kohlbrecher00}
\bibinfo{author}{\bibfnamefont{J.}~\bibnamefont{Kohlbrecher}} \bibnamefont{and}
  \bibinfo{author}{\bibfnamefont{W.}~\bibnamefont{Wagner}},
  \bibinfo{journal}{J. Appl. Cryst.} \textbf{\bibinfo{volume}{33}},
  \bibinfo{pages}{804} (\bibinfo{year}{2000}).

\bibitem[{\citenamefont{Keiderling}(2002)}]{Keiderling02}
\bibinfo{author}{\bibfnamefont{U.}~\bibnamefont{Keiderling}},
  \bibinfo{journal}{Appl. Phys. A} \textbf{\bibinfo{volume}{74}},
  \bibinfo{pages}{S1455} (\bibinfo{year}{2002}).

\bibitem[{\citenamefont{Wilhelm et~al.}(1998)\citenamefont{Wilhelm, Maring, and
  Spiess}}]{Wilhelm98}
\bibinfo{author}{\bibfnamefont{M.}~\bibnamefont{Wilhelm}},
  \bibinfo{author}{\bibfnamefont{D.}~\bibnamefont{Maring}}, \bibnamefont{and}
  \bibinfo{author}{\bibfnamefont{H.}~\bibnamefont{Spiess}},
  \bibinfo{journal}{Rheol. Acta} \textbf{\bibinfo{volume}{37}},
  \bibinfo{pages}{399} (\bibinfo{year}{1998}).

\bibitem[{\citenamefont{Salmon et~al.}(2003)\citenamefont{Salmon, Colin, and
  Manneville}}]{Salmon03c}
\bibinfo{author}{\bibfnamefont{J.-B.} \bibnamefont{Salmon}},
  \bibinfo{author}{\bibfnamefont{A.}~\bibnamefont{Colin}}, \bibnamefont{and}
  \bibinfo{author}{\bibfnamefont{S.}~\bibnamefont{Manneville}},
  \bibinfo{journal}{Phys. Rev. Lett.} \textbf{\bibinfo{volume}{90}},
  \bibinfo{pages}{228303} (\bibinfo{year}{2003}).

\bibitem[{\citenamefont{Dhont and Briels}(2008)}]{dhont08}
\bibinfo{author}{\bibfnamefont{J.~K.~G.} \bibnamefont{Dhont}} \bibnamefont{and}
  \bibinfo{author}{\bibfnamefont{W.~J.} \bibnamefont{Briels}},
  \bibinfo{journal}{Rheol. Acta} \textbf{\bibinfo{volume}{47}},
  \bibinfo{pages}{257} (\bibinfo{year}{2008}).

\bibitem[{\citenamefont{Olmsted}(2008)}]{olmsted08}
\bibinfo{author}{\bibfnamefont{P.~D.} \bibnamefont{Olmsted}},
  \bibinfo{journal}{Rheol. Acta} \textbf{\bibinfo{volume}{47}},
  \bibinfo{pages}{283} (\bibinfo{year}{2008}).

\bibitem[{\citenamefont{Lettinga and Dhont}(2004)}]{Lettinga04a}
\bibinfo{author}{\bibfnamefont{M.~P.} \bibnamefont{Lettinga}} \bibnamefont{and}
  \bibinfo{author}{\bibfnamefont{J.~K.~G.} \bibnamefont{Dhont}},
  \bibinfo{journal}{J. Phys.: Condens. Matter} \textbf{\bibinfo{volume}{16}},
  \bibinfo{pages}{S3929} (\bibinfo{year}{2004}).

\bibitem[{\citenamefont{Chen}(1993)}]{chen93}
\bibinfo{author}{\bibfnamefont{Z.~Y.} \bibnamefont{Chen}},
  \bibinfo{journal}{Macromolecules} \textbf{\bibinfo{volume}{26}},
  \bibinfo{pages}{3419} (\bibinfo{year}{1993}).

\bibitem[{\citenamefont{Picken et~al.}(1990)\citenamefont{Picken, Aerts,
  Visser, and Northolt}}]{Picken90}
\bibinfo{author}{\bibfnamefont{S.~J.} \bibnamefont{Picken}},
  \bibinfo{author}{\bibfnamefont{J.}~\bibnamefont{Aerts}},
  \bibinfo{author}{\bibfnamefont{R.}~\bibnamefont{Visser}}, \bibnamefont{and}
  \bibinfo{author}{\bibfnamefont{M.~G.} \bibnamefont{Northolt}},
  \bibinfo{journal}{Macromolecules} \textbf{\bibinfo{volume}{23}},
  \bibinfo{pages}{3849} (\bibinfo{year}{1990}).

\bibitem[{\citenamefont{Tirado et~al.}(1980)\citenamefont{Tirado, Martinez, and
  Delatorre}}]{Tirado80}
\bibinfo{author}{\bibfnamefont{M.}~\bibnamefont{Tirado}},
  \bibinfo{author}{\bibfnamefont{C.}~\bibnamefont{Martinez}}, \bibnamefont{and}
  \bibinfo{author}{\bibfnamefont{J.~G.} \bibnamefont{Delatorre}},
  \bibinfo{journal}{J. Chem . Phys.} \textbf{\bibinfo{volume}{73}},
  \bibinfo{pages}{1986} (\bibinfo{year}{1980}).

\bibitem[{\citenamefont{Tao et~al.}(2006)\citenamefont{Tao, Dhont, and
  Briels}}]{Tao06c}
\bibinfo{author}{\bibfnamefont{Y.-G.} \bibnamefont{Tao}},
  \bibinfo{author}{\bibfnamefont{J.~K.~G.} \bibnamefont{Dhont}},
  \bibnamefont{and} \bibinfo{author}{\bibfnamefont{W.~J.}
  \bibnamefont{Briels}}, \bibinfo{journal}{J. Chem. Phys.}
  \textbf{\bibinfo{volume}{124}},
  \bibinfo{pages}{134906} (\bibinfo{year}{2006}).

\end{thebibliography}
\end{document}